\documentclass[twocolumn,showpacs]{revtex4}

\usepackage{amsmath}
\usepackage{amssymb}
\usepackage{epsfig}

\newcommand{\beq}{\begin{equation}}
\newcommand{\eeq}{\end{equation}\noindent}
\newcommand{\bean}{\begin{eqnarray*}}
\newcommand{\eean}{\end{eqnarray*}\noindent}
\newcommand{\bea}{\begin{eqnarray}}
\newcommand{\eea}{\end{eqnarray}\noindent}

\begin{document}

\title{Neutrino-nucleus interactions as a probe to constrain 
double-beta decay predictions}
\author{Cristina Volpe}
\email{volpe@ipno.in2p3.fr}
\affiliation{Groupe de Physique Th\'{e}orique, Institut de Physique Nucl\'{e}aire,
F-91406 Orsay Cedex, France}

\date{\today}

\begin{abstract}
We propose to use charged-current neutrino-nucleus interactions as a probe 
of the many virtual
transitions involved in neutrinoless double-beta decay.
By performing $\nu$ and $\bar{\nu}$
interaction studies on the initial  and final nucleus respectively,
one can get information
on the two branches involved in
the double-beta decay process.
The measurement of such reactions could help to further constrain
double-beta decay predictions on the half-lives.
\end{abstract}

\maketitle
The experimental evidence of neutrino oscillations has opened a new
exciting era in neutrino physics \cite{osc}. The observation of non-zero
neutrino masses and mixings is important for various domains of physics,
from high-energy physics to cosmology, and has triggered 
many theoretical investigations as well as experimental
projects. This discovery represents
a big step forward in our knowledge of neutrinos properties. Still, 
these particles have many mysteries to unveil,
one of the key open questions being 
the Dirac or Majorana nature of neutrinos, which  
is currently explored with the search for neutrinoless double-beta decay
in nuclei. The observation of such a process would represent an outstanding
discovery. Besides, important information on the neutrino mass scale, on the
hierarchy and on the Majorana phases could be obtained (see e.g. \cite{info}).

In the last decades important progress has been done on the double-beta 
decay process, both from the experimental 
and from the theoretical points of view
\cite{rev1,suhoreport,rev2,rev3,rev4,rev5,rev6}.
Concerning the experimental achievements,
the two-neutrino ($2 \nu$)
double-beta decay process is now observed in
a large ensemble of nuclei, and the half-lives for several
isotopes are being measured 
with a very high precision \cite{nemo3}.
As far as the neutrinoless ($0 \nu$) 
double-beta decay is concerned,
the presently  running NEMO3 experiment
 will reach a sensitivity of 0.2 eV in the near future \cite{nemo3}.
The best present limit, in the range of
0.3-1.eV, comes from the germanium-based experiments \cite{ge}. 
In particular, 
a claim for the evidence of neutrinoless double-beta decay 
was made recently  \cite{evidence}. This will be
confirmed or refuted by 
the future CUORE \cite{cuore} and GERDA
experiments. 
Various other 
projects  are now under study to reach the tens of meV
sensitivity \cite{rev5,rev6,zdesenko}.
From the theoretical point of view, present
predictions of neutrinoless double-beta decay half-lives 
still exhibit important differences for the same candidate emitter. 
Understanding the origin of the differences
in the nuclear matrix elements that determine such decay rates
represents an essential step in the future, 
both for the search of neutrinoless double-beta
decay, and to fully exploit a positive result.

So far, several related processes have been considered
to constrain theoretical calculations:
 the
charge-exchange reactions \cite{charge-exchange},
$\beta_+$ and $\beta_-$ decays 
(see e.g.\cite{suhoreport,betadecay}) 
and muon capture \cite{muon-capture}.
Different states are probed in  
these processes. 
In charge-exchange reactions mainly information on the
allowed Gamow-Teller transitions is obtained.
Concerning beta-decay, only transitions to low-energy
states are probed, whereas 
high-energy states can be 
excited in muon capture, which 
explores one of the two branches
of the double-beta decay process only.

In this letter, we propose to use charged-current neutrino-nucleus 
interactions 
as a supplementary probe to constrain the calculations
on the neutrinoless double-beta decay half-lives.
We discuss
that such processes probe the many virtual states involved.
In particular, neutrino-nucleus interactions have 
two specific features compared to the double-beta
decay related processes considered until now.
If both $\nu$ and $\bar{\nu}$ beams are available,
the measurement of $\nu$ ($\bar{\nu}$) 
interactions on the initial (final)
nucleus can probe the two branches. 
The relevant information on the 
Fermi and Gamow-Teller type (also called forbidden)
transitions could be gathered by varying the impinging neutrino
energy, since the contribution of these states 
to the total cross section
increases for increasing neutrino energy.
We illustrate this point by taking $^{48}$Ca as a typical
example.
 Such measurements could be performed at 
facilities producing low-energy neutrinos, 
like the one proposed at SNS and exploiting
conventional sources (pion and muon decay-in-flight 
or decay-at-rest)  
\cite{orland},
or with a low-energy beta-beam facility 
\cite{lownu} which uses the beta-decay of boosted
radioactive ions \cite{zucchelli}. 
The latter 
option has two particularly attractive features.
First, very pure $\nu_e$ 
and $\bar{\nu}_e$ beams can be produced and, second,
the neutrino average energy can be easily varied by modifying the
acceleration of the ions (note that $E_{\nu}=2 \gamma Q_{\beta}$ where
$\gamma$ and $Q_{\beta}$ are the Lorentz factor and the beta Q-value
respectively). The interaction rates are discussed in \cite{serreau,mclaughlin}.

The transition operators involved in the neutrinoless double-beta decay
are given in detail in e.g. \cite{rev1,rev2,suhoreport}.
These have two-body character and involve spin-isospin and isospin
degrees of freedom. In the $2 \nu$ case, the energy involved is
typically a few MeV, like in beta-decay. Therefore, mainly allowed
Gamow-Teller transitions are involved. The measured
$2 \nu$ half-lives are often used to constrain the calculations
and, in particular, to determine the intensity of the particle-particle
interaction (see e.g. the discussion in \cite{2beta} and references
therein).
On the other hand, in the $0 \nu$ case due, 
for example, to a massive Majorana neutrino
exchange, from the uncertainty principle one gets that
the typical momentum is expected to be 
of the order of 100 MeV.
Therefore, both Fermi and Gamow-Teller -- 
allowed and forbidden -- transitions
from the initial (and final) nucleus to the intermediate one can be excited,
arising from
 the multipole expansion of the neutrino-exchange
potential (see e.g.\cite{multipole}).
The contribution of these states to the half-lives of various
double-beta decay emitters has been explicitly shown in several
works, 
like for the case of $^{48}$Ca (see Figs. 5 and 6 of Ref. \cite{suhoca}
and Fig.3 of Ref.\cite{multipole}) and for the case of $^{76}$Ge
(see Figs.2 of Refs.\cite{src,geklap} and
Fig.1 of \cite{tomoda}). In particular, Ref.\cite{src} discusses
the effect of short-range correlations on the contribution of
these states. Using the multipole expansion of the neutrino potential,
it is easy to show that the Fermi and Gamow-Teller type transitions
are the same as the ones involved in neutrino-nucleus interactions.
In fact, 
the general expression for the (anti)neutrino-nucleus reaction cross section
can be written as \cite{kubo}:
\begin{equation}
\sigma={G^{2} \over {2 \pi}}cos^{2}\theta_C\sum_{f}p_lE_l
\int_{-1}^{1}d(cos \, \theta)M_{\beta},
\label{e:2}
\end{equation}
where $G \,cos \, \theta_C$ is the weak coupling constant, 
$E_l=E_{\nu}-E_{f}+E_{i}$ ($p_{l}$) is the outgoing lepton
energy (momentum), $E_{f}$ ($E_{i}$) and $E_{\nu}$ 
being the energy of the final (initial) nuclear
state and  the incident neutrino energy respectively and
$\theta$ is
the angle between the directions of the incident neutrino and the outgoing
lepton. The transition matrix element $M_{\beta}$ is given by:
\begin{equation}
M_{\beta} \equiv
M_{F} \vert  \langle f \vert \tilde 1 \vert i \rangle \vert^2 +
M_{G0} {1 \over 3}
\vert  \langle f \vert \tilde \sigma \vert i \rangle \vert^2 +
M_{G2} \Lambda
\label{e:3}
\end{equation}
where, in particular,  
the squared nuclear matrix elements involve isospin and
spin-isospin degrees of freedom and a generalized Bessel 
function as a radial part:
\begin{equation}
\vert  \langle f \vert \tilde 1 \vert i \rangle \vert^2=
a_i \sum_l \vert \langle {\it J_f \Vert
\sum_{k} t_{+}(k) j_l(qr_k)Y_l(\hat{{\bold r}}_k) \Vert J_i } \rangle
\vert^2,
\label{e:4}
\end{equation}
  
\begin{equation}
\vert  \langle f \vert \tilde \sigma \vert i \rangle \vert^2=
a_i \sum_{l,K} \vert \langle {\it J_f \Vert
\sum_{k} t_{+}(k) j_l(qr_k)[Y_l(\hat{{\bold r}}_k)
\times {\bold \sigma}]^{(K)} \Vert J_i } \rangle
\vert^2.
\label{e:5}
\end{equation}
\noindent
Here, $k$ labels the space and spin-isospin coordinates of the
$k$-th nucleon,
$l,~l'$ are the orbital angular momenta and $K$ is the total
angular momentum of
the transition operators and with 
$a_i= 4 \pi/(2 J_i +1)$, $b_{l,l'}=(-1)^{l/2-l'/2+K} \sqrt{2l+1} \sqrt{2l'+1}$.
The coefficients $M_{F}, M_{G0}$ and $M_{G2}$ \cite{kubo}
appearing in (\ref{e:3})
depend on the momentum $q$ transferred to the nucleus
and on the standard  
form factors.
(Note that, since neutrino-nucleus interactions are essentially 
a one-body process,
the cross sections are not influenced by short-range correlations.)
In the limit of small momentum transfer, 
the operators in (\ref{e:4}-\ref{e:5}) reduce to the allowed Fermi and Gamow-Teller
operators. On the other hand the $q$-dependent operators are the same as the
ones involved in muon capture (except for the radial dependence which, in
that case, includes the muon wave function). 

\begin{figure}[t]
\includegraphics[angle=-90.,width=6cm]{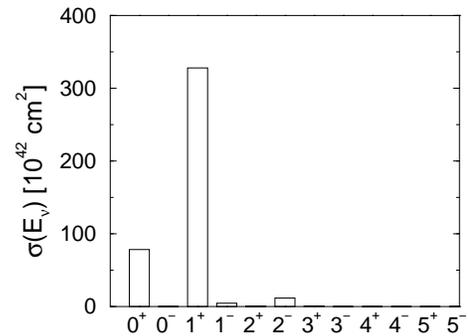}
\caption{Contribution of the states of different multipolarity 
to the total charged-current  $\nu_e + ^{48}$Ca 
cross section for neutrino energy $E_{\nu}=30~$MeV.
The histograms show the contribution of the Fermi 
($J^{\pi}=0^+$), the Gamow-Teller  ($1^+$) and the spin-dipole 
($0^-,1^-,2^-$) states and all higher multipoles up to $5$.
\label{fig:1}}
\end{figure}

\begin{figure}[t]
\includegraphics[angle=-90.,width=6cm]{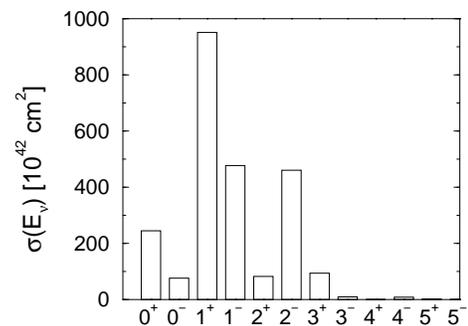}
\caption{Same as Fig.1 for $E_{\nu}=60~$MeV.  \label{fig:2}}
\end{figure}

The possible future availability of neutrino beams
having several tens of MeV could offer the opportunity to perform
neutrino-nucleus interaction studies at different energies,
exploiting either the decay-at-rest or in flight of muons and
pions \cite{orland}, or low-energy beta-beams \cite{lownu}.
The relevant information on the transition matrix elements
could be disentangled by varying the impinging neutrino energy
since the importance of these states -- relative to the allowed
Fermi and Gamow-Teller ones -- varies with the neutrino energy.
We illustrate this feature by taking the case of $^{48}$Ca as
typical example.
The results are obtained by using the microscopic proton-neutron 
Quasi-Particle
Random Phase Approximation (QRPA) using Skyrme forces.
The details of the approach can be
found in \cite{c12}.
The pairing gap is 
taken to reproduce experimental separation energies
and the force used is SGII. Similar trends
are obtained with the SIII force.
Note that, 
as far as  the double-beta decay of $^{48}$Ca is concerned, several
calculations exist in the literature either
within the shell model approach \cite{rev1,ca48sm} or within the QRPA approach
and its variants \cite{suhoca,multipole}.

Figures 1-4 illustrates how 
the contribution of states of different multipolarity
to the total $\nu_e + ^{48}$Ca and $\bar{\nu}_e + ^{48}$Ti cross 
sections
evolves for increasing 
neutrino energies, i.e. $E_{\nu,\bar{\nu}}$=30 and 60 MeV.
Similar results can be obtained for other candidate
double-beta decay emitters.
From Figs. 1 and 3 one can see that the cross 
section at $E_{\nu,\bar{\nu}}=30$~MeV
are mainly dominated by the allowed Fermi and 
the Gamow-Teller states,  even
though there can be a significant contribution coming from the other multipoles
(Figure 3). 
The situation is quite different for $E_{\nu,\bar{\nu}}=60$~MeV where
the contribution of the spin-dipole ($J^{\pi}=0^-,1^-,2^-$) states as well as
higher multipoles (mainly the $J^{\pi}=2^+$ and $3^+$ in this case) 
becomes as important (or even dominates) the total cross section. 
Note that such trend can appear already at lower
neutrino energies for heavier nuclei.
These results
indicate that neutrino beams in the 100 MeV energy range 
could help improving our knowledge of
these states and furnish 
a supplementary constrain, through neutrino-nucleus measurements, 
on the matrix elements
involved in $0 \nu$ double-beta decay. 
\begin{figure}[t]
\includegraphics[angle=-90.,width=6cm]{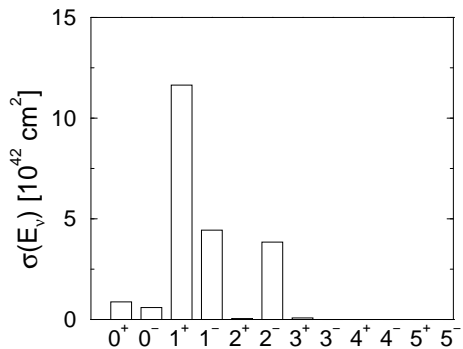}
\caption{Same as Fig.1 for $\bar{\nu} + ^{48}$Ti. 
\label{fig:3}}
\end{figure}

\begin{figure}[t]
\includegraphics[angle=-90.,width=6cm]{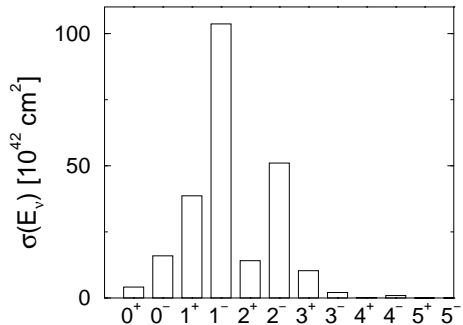}
\caption{ Same as Fig.3 for $E_{\nu}=60~$MeV.  \label{fig:4}}
\end{figure}

In conclusion, the search for neutrinoless double-beta decay is a crucial issue
to answer the question of the Dirac or of the Majorana nature of neutrinos.
The theoretical predictions on the half-lives still suffer from important 
differences
for the same double-beta decay emitter. So far various processes have been considered
to constrain the calculations. Here we have argued that
the forbidden
transitions which contribute to the neutrinoless half-lives in the case, e.g.
of a massive Majorana neutrino exchange, are the same as the ones involved in
neutrino-nucleus interactions.
We have also discussed that the information on such states might be
extracted if neutrino beams with different energies are available,
since their role increases for increasing neutrino energy.
The study of such processes has the
advantage of offering the possibility, at least in principle, 
to explore the two branches,
through 
the interaction of (anti-)neutrinos from the initial (final)
nucleus to the intermediate one.  
We have illustrated these features for a typical example. 
Future facilities producing low-energy neutrino beams
can offer the opportunity to perform such studies.
We hope that this work will trigger studies on 
the feasibility of such experiments.

\vspace{.2cm}
We thank F. Simkovic for useful discussions, R. Lombard and J. Serreau for
careful reading of the manuscript.


\begin{thebibliography}{99}
\bibitem{osc} S.M. Bilenky, C. Giunti, J.A. Grifols, E. Masso,
Phys. Rept. 379, 69 (2003) and references therein.

\bibitem{info}  S.M. Bilenky, A. Faessler, F. Simkovic,
Phys. Rev. {\bf D} 70, 033003 (2004);
S. Pascoli, S.T. Petcov, W. Rodejohann, Phys. 
Lett. B 558, 141 (2003); 
H.V. Klapdor-Kleingrothaus, H. P\"as, A.Y. Smirnov,
Phys. Rev. D 63, 073005 (2001).

\bibitem{rev1} W.C. Haxton and G.F. Stevenson, 
Prog. Part. Nucl. Phys. {\bf 12}, 409 (1984).
\bibitem{suhoreport}
J. Suhonen and O. Civitarese, Phys. Rep. 300, 123 (1998).
\bibitem{rev2} A. Faessler and F. Simkovic,
J. Phys. G: Nucl. Part. Phys. {\bf 24}, 2139 (1998).
\bibitem{rev3} J.D. Vergados, Phys. Atom. Nucl. {\bf 63}, 1137
(2000).
\bibitem{rev4} H.V. Klapdor-Kleingrothaus, ``Sixty years of double 
beta decay: from nuclear
physics to beyond the standard model'', 
Singapore: World Scientific (2001).
\bibitem{rev5} 
S.R. Elliott and P. Vogel,  Ann. Rev. Nucl. Part. Sci. 52, 115 (2002).
\bibitem{rev6} S.R. Elliott and 
J. Engel, J. Phys. G 30, R183 (2004).

\bibitem{nemo3} 
X. Sarazin, Proceedings to ``Neutrino 2004'' [hep-ex/0412012].


\bibitem{ge} L. Baudis {\it et al.}, Phys. Rev. Lett.
{\bf 83}, 41 (1999);
H.V. Klapdor-Kleingrothaus {\it et al.}, 
Eur. Phys. J. A 12 (2001) 147; C.E. Aalseth,
Phys. Rev. D 65, (2002). 

\bibitem{evidence} 
H.V. Klapdor-Kleingrothaus {\it et al.}, 
Phys. Lett. B 578, 54 (2004).


\bibitem{cuore} The CUORE Collaboration, Nucl. 
Instrum. Meth. A 518, 775 (2004).

\bibitem{zdesenko} 
Yu. Zdesenko, Rev. Mod. Phys. 74, 663 (2003). 

\bibitem{charge-exchange} H. Akimune {et al.}, Phys. Lett. B 394, 23 (1997);
J. Bernabeu {et al.}, IPNO/TH-88-58, FTUV-88/20.

\bibitem{betadecay} K. Muto, E. Bender and H.V. Klapdor,
Z. Phys. A {\bf 334}, 177 (1989); M. Aunola, J. Suhonen,
Nucl. Phys. A {\bf 602}, 133 (1996).

\bibitem{muon-capture}  M. Kortelainen and J. Suhonen,
Phys. Atom. Nucl. 67, 1202 (2004); M. Kortelainen and J. Suhonen,
Europhys. Lett. 58, 666 (2002).


\bibitem{orland} F.T. Avignone {\it et al.},
Phys. Atom. Nucl.63, 1007 (2000); see http://www.phy.ornl.gov/orland/;
Y. Efremenko, private communication.

\bibitem{lownu} C. Volpe, Jour. Phys. G {\bf 30}, L1 (2004)
[hep-ph/0303222].

\bibitem{zucchelli} P. Zucchelli, Phys. Lett. B {\bf 532}, 166 (2002).

\bibitem{serreau} J. Serreau and C. Volpe, Phys. Rev. C {\bf 70},
055502 (2004).

\bibitem{mclaughlin} G.C. McLaughlin, Phys. Rev. C 70, 045804 (2004) 

\bibitem{2beta} V.A. Rodin {\it et al.}, Phys. Rev. C {\bf 68},
044302 (2003) and references therein.

\bibitem{multipole} K. Muto, E. Bender and H.V. Klapdor-Kleingrothaus,
Z. Phys. A {\bf 339}, 435 (1991).

\bibitem{suhoca} J. Suhonen, J. Phys. G {\bf 19}, 139 (1993).

\bibitem{src} A. Bobyk, W.A. Kaminski, F. Simkovic,
Phys. Rev. C 63, 051301 (2001). 

\bibitem{geklap} K. Muto, E. Bender and H.V. Klapdor-Kleingrothaus,
Z. Phys. A {\bf 334}, 187 (1989).

\bibitem{tomoda} T. Tomoda and A. Faessler, Phys. Lett. B {\bf 199},
475 (1987).

\bibitem{kubo} T. Kuramoto, M. Fukugita, Y. Kohyama and K. Kubodera,
Nucl. Phys. A 512, 711 (1990).

\bibitem{c12} C. Volpe {\it et al.}, Phys. Rev. C {\bf 62}, 015501 (2000). 

\bibitem{ca48sm} J. Retamosa, E. Caurier, F. Nowacki, Phys. Rev. C
{\bf 51}, 371 (1995); E. Caurier, A.P. Zuker, A. Poves,
Phys. Lett. B {\bf 252}, 13 (1990).


\end{thebibliography}
\end{document}